\documentclass[runningheads]{svmult}
\usepackage{makeidx}   
\usepackage{graphicx}  
\usepackage{subeqnar}  
\usepackage{multicol}  
\usepackage{cropmark} 
\usepackage{physprbb}  
\newcommand{\greeksym}[1]{{\usefont{U}{psy}{m}{n}#1}}
\newcommand{\umu}{\mbox{\greeksym{m}}}

\def\ra{\rightarrow}
\def\be{\begin{equation}}
\def\ee{\end{equation}}
\def\bea{\begin{eqnarray}}
\def\eea{\end{eqnarray}}
\newcommand{\expe}{experiment}
\def\Journal#1#2#3#4{{#1} {\bf #2}, #3 (#4)}

\def\NIMA{{\em Nucl. Instrum. Methods} A}

\def\NPBP{{\em Nucl. Phys.} B (Proc. Suppl.)}

\def\PLB{{\em Phys. Lett.}  B}
\def\PRL{\em Phys. Rev. Lett.}
\def\PRD{{\em Phys. Rev.} D}

\def\PRP{{\em Phys. Rep. }}

\def\PNPP{\em Prog. Nucl. Part. Phys.}

\def\PNPP{\em Prog. Nucl. Part. Phys.}

\newcommand{\exps}{experiments }
\newcommand{\evs}{\mbox{eV$^2$} }
\newcommand{\app}{appearance }
\newcommand{\dis}{disappearance }

\newcommand{\osz}{oscillation }
\newcommand{\oszs}{oscillations }

\newcommand{\lbls}{long baseline experiments }

\newcommand{\delm}{\mbox{$\Delta m^2$} }
\newcommand{\bnel}{\mbox{$\bar{\nu}_e$} }
\newcommand{\bnmu}{\mbox{$\bar{\nu}_\mu$} }

\newcommand{\nel}{\mbox{$\nu_e$} }
\newcommand{\nmu}{\mbox{$\nu_\mu$} }
\newcommand{\ntau}{\mbox{$\nu_\tau$} }
\newcommand{\sint}{\mbox{$\sin^2 2\theta$} }
\newcommand{\sk}{Super-Kamiokande }
\newcommand{\neu}{neutrino }
\newcommand{\neus}{neutrinos }
\newcommand{\cms}{\mbox{$cm^{-2}s^{-1}$} }

\newcommand{\nmune}{\mbox{$\nu_{\mu} - \nu_e$} }
\newcommand{\nmuntau}{\mbox{$\nu_{\mu} - \nu_{\tau}$} }

\begin{document}
\title*{Long baseline neutrino oscillation experiments}
%
%
\toctitle{Long baseline neutrino oscillation experiments}
%
%
%
\author{Kai Zuber\inst{}}
%
%
%
\institute{Lehrstuhl f\"ur Exp. Physik IV, Universit\"at Dortmund,
Otto-Hahn Str. 4, 44221 Dortmund, Germany
\footnote{Current address: Department of Physics, University of
Oxford, Denys Wilkinson Building, Keble Road,
Oxford OX1 3RH, England}
}

\maketitle              

\begin{abstract}
Evidences for neutrino oscillations coming from atmospheric and solar observations
can be probed by terrestrial long baseline experiments. This requires accelerator
beams or nuclear power plants. The current status of these searches as well as future
activities are discussed. A precise determination of all matrix elements and searches
for a leptonic CP-violation will be possible with high intensity accelerator beams
currently discussed in the context of neutrino factories.
\index{abstract} 
\end{abstract}

\section{Introduction}
\label{sect1}
A non-vanishing rest mass ot the \neu{} has far reaching consequences from cosmology down
to particle
physics. For recent reviews see \cite{kai,samoil}.
In the last years growing evidence for such a mass arose in \neu \osz \exps .
In a simple two flavour mixing scheme the \osz{} probability $P$ is given by 
\be
P (L/E) = \sint{} sin^2(1.27 \delm_{ij} L/E)
\ee
with \delm$_{ij}$ = $\mid m_j^2 -m_i^2 \mid$, L the source-detector distance and E the
\neu{} energy.
The first evidence comes from
the LSND-experiment \cite{lsnd}, observing an effect in the \bnmu - \bnel channel. However a large
fraction of the possible parameters are in contradiction with other \exps especially KARMEN \cite{eitel}
and also for \delm $>$ 10 \evs with NOMAD \cite{slava}. The most likely parameter sets are
\delm $\approx 1$ \evs and mixing angles about \sint $\approx$ 10$^{-3}$, see also the results
from a combined data analysis \cite{eitel2}. 
The second evidence is coming from the zenith
angle distribution of atmospheric \neus as observed by \sk , showing a clear deficit in upward going \nmu \neus \cite{skevi} meanwhile
confirmed by other experiments \cite{macro,soudan2} .
For \sk it  can be explained by \oszs with \delm in the range $1.6 - 4 \cdot 10^{-3}$ \evs and \sint $>$ 0.89 (90 \% CL)
assuming \nmu -\ntau transitions.
Last not least there is
the long standing solar \neu{} problem, which got some major new impacts during the last two years.
By comparing neutrino-electron scattering rates from \sk and first charged current results relying
on inverse beta-decay from SNO it already became clear that active \neu flavours besides \nel are coming
from the sun \cite{snocc}. With the recent results on neutral current data and day/night effects in CC events,
SNO was able to show, that indeed the dominant part of the solar neutrino flux is coming in a non \nel
flavour but the total flux is in agreement with standard solar model predictions \cite{snonc}. Furthermore, to describe the data,
the large mixing angle solution (LMA) is now in strong favour, having a best fit value of \delm = 5 $\cdot 10^{-5}$ \evs and a mixing 
of $\tan^2 \theta $ = 0.34 \cite{snoall}.\\
With these evidences in hand two goals emerge which can be explored by long baseline experiments, 
meaning a large distance of the detector from the neutrino source. First of all to prove
the atmospheric and solar evidence and disentangle the \osz channel using neutrinos from accelerators or nuclear
power plants. The second goal is the fact, that we have to face a 3$\times$3 mixing matrix among the
neutrinos called MNS-matrix, in analogy to the CKM-matrix in the quark sector. 
It can be written in a convenient way as
\be
\left(\begin{array}{ccc}
\cos_{12} & \sin_{12} & 0 \\
-\sin_{12} & \cos_{12} & 0\\
0 & 0 & 1 \\
\end{array}\right)    
\left(\begin{array}{ccc}
\cos_{13} & 0 & \sin_{13} \\
0 & 1 & 0 \\
-\sin_{13}e^{i\delta} & 0 & \cos_{13}e^{i\delta} \\
\end{array}\right)  
\left(\begin{array}{ccc}
1 & 0 & 0 \\
0 & \cos_{23} & \sin_{23} \\
0 & -\sin_{23} & \cos_{23} \\
\end{array}\right)   
\ee
Ignoring the LSND result the first part could describe solar neutrinos, while the third one corresponds to 
atmospheric neutrinos. However, as can be seen from the part in the middle also CP-violation could in principle occur, requiring 
$\sin \theta_{13}$ to be non-zero. Therefore as a long term perspective, the measurements of all matrix elements together
with a possible CP-violation and matter effects, which are at work in the Sun, could be explored. 
This is the goal for what is generally called ''neutrino factories''.

\section{Reactor experiments}
\label{sect2}
\begin{figure}[b]
\begin{center}
\includegraphics[width=5cm,height=5.5cm]{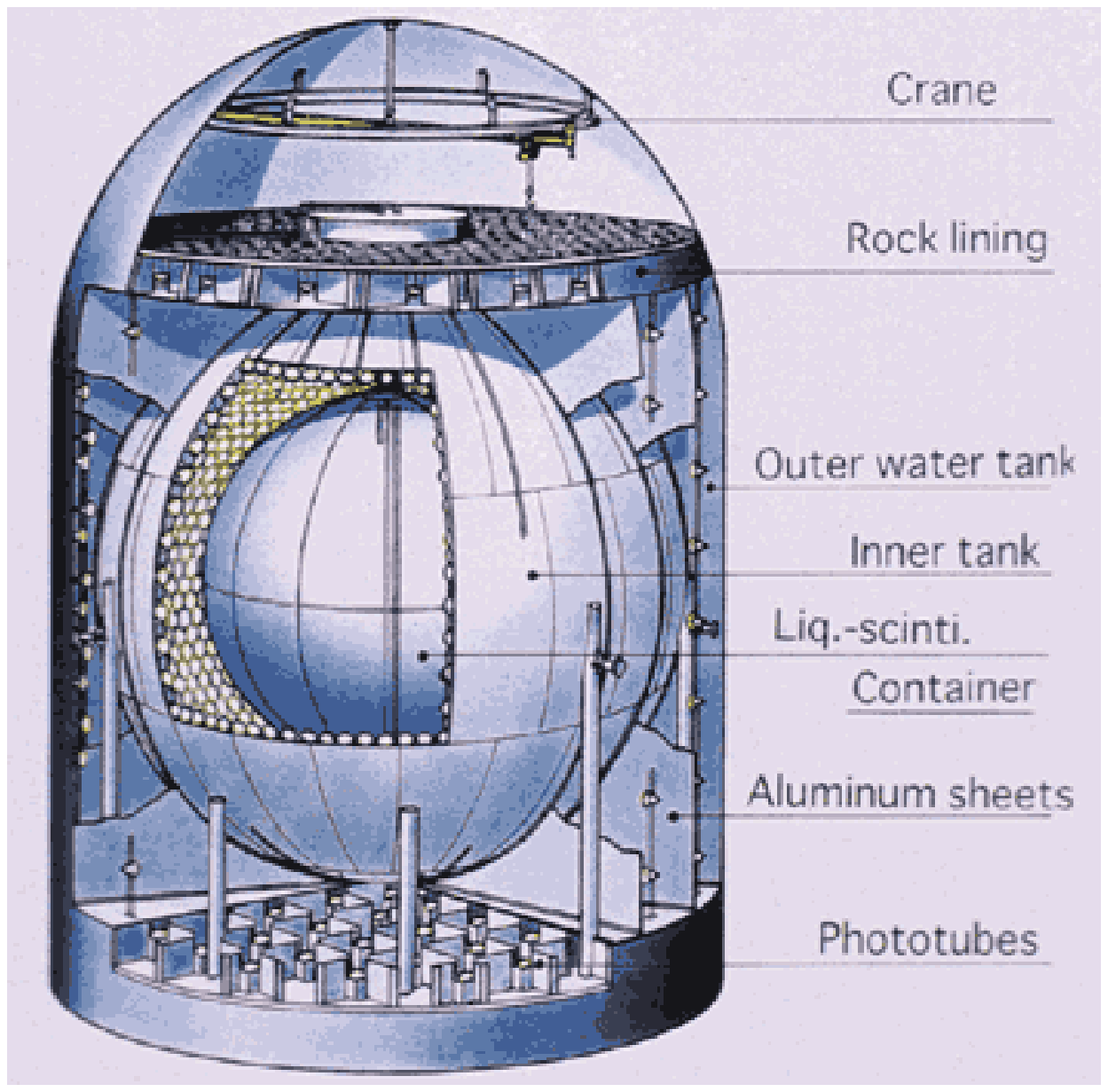}
\includegraphics[width=5cm,height=5.5cm]{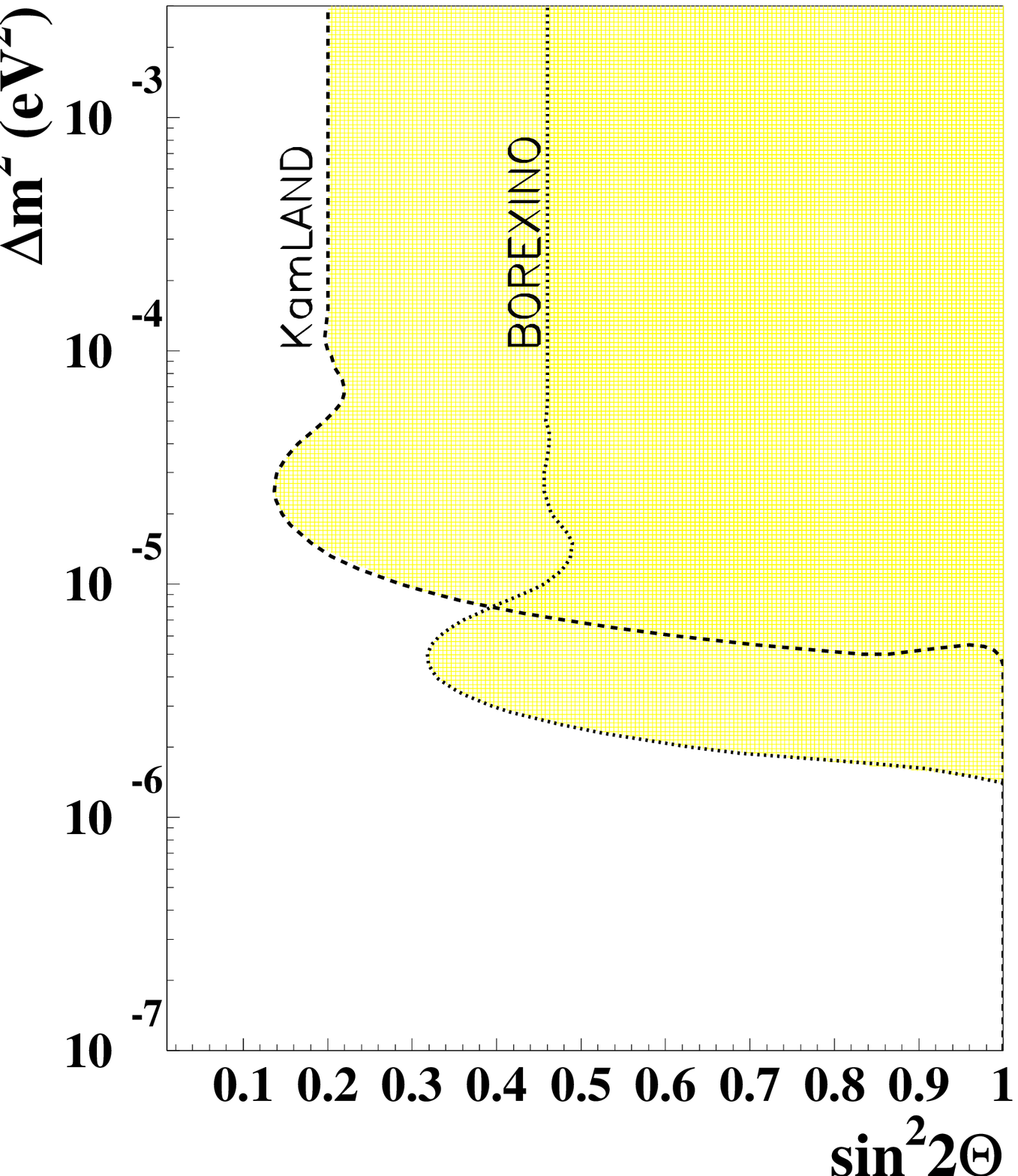}
\end{center}
\caption[]{Left: Schematic view of the KamLAND - detector, which started data taking recently. Right: Expected sensitivity curves
for KamLAND and Borexino. The best fit value of the LMA solution is \delm = $5 \cdot 10^{-5}$ \evs (from \protect \cite{stefan}).}
\label{fig1}
\end{figure}

Reactor experiments are disappearance
experiments looking for \bnel $\ra \bar {\nu}_X$.
Reactors are a source of MeV \bnel due to the 
fission products being $\beta$-unstable. 
Experiments typically
try to measure the positron spectrum, which can be deduced from the \bnel -
spectrum, and either compare it directly to the theoretical predictions
or measure it at several distances from the reactor and search for spectral
changes.
Both types of experiments were done in the past. 
The detection reaction is
\be
\label{gl1}
\bnel + p \ra e^+ + n
\ee
with an energy threshold of 1.804 MeV.
Coincidence techniques are used between the
annihilation
photons and the neutrons
which diffuse and
thermalize within 10-100 $\mu$s. The reactions commonly used for neutron detection
are $p(n,\gamma)D
$ and $Gd(n,\gamma)Gd$ resulting
in gamma rays of either 2.2 MeV or up to 8 MeV.
The main background are cosmic ray muons producing neutrons
in the surrounding of the detector.\\
With respect to past reactor searches, the recent \exps{} CHOOZ and Palo Verde can already be
considered as 
\lbls{}. Their distance to the power stations of 1030 m and about 800 m respectively is 
already a factor of at
least three larger than any other reactor \expe{} done before. The results from CHOOZ \cite{apo98}
and Palo Verde \cite{paloverde}
already exclude \nmune{} \oszs{} as explanation for the atmospheric \neu anomaly. Long-baseline \exps{} 
even by accelerator
definitions will be done by KamLAND and BOREXINO.\\
The KamLAND \expe{} \cite{kamland} is installed in the Kamioka mine in Japan
(Fig.~\ref{fig1}). It contains
1000t of Liquid
Scintillator as a
main target, filled in a plastic balloon. Outside the balloon within a stainless steel sphere of 18 m diameter
there is a mixture of paraffin oils. Together they form the inner detector.
The readout is done with 1878 photomultipliers. 
The steel sphere is surrounded by buffer water with
a total mass of 2500t. 
There are 6 reactors with a total thermal power of 69 GW in a distance
between 140 km and 210 km to
Kamioka which act as dominant \bnel -source. They produce a total \neu flux of $1 \cdot 10^6$
\cms{} at Kamioka for antineutrino energies larger than 1.8 MeV, resulting in 2 events/day.
This will allow to measure \delm{} below $10^{-5} eV^2$,
therefore probing
the LMA solution of the solar \neu{} problem. If the background can be
reduced by another factor of ten
with respect to the proposed value, even the direct observation of solar $^{7}$Be and
terrestrial \neus{} seems
feasible. Data taking started in January 2002.\\
Originally proposed for solar \neu detection, also the BOREXINO \expe{}
has the ability to investigate
reactor
\neus{} \cite{stefan}. The \bnel-flux at Gran Sasso Laboratory is around $1.5 \cdot 10^5$ \cms{}
for energies larger
than 1.8 MeV produced by power plants typically 800 km away. Without \osz{} this
would result in 27
events/year in a 300 t liquid scintillation detector. The sensitivity might go
down to \delm $\approx 10^{-6} eV^2$ and \sint $>$ 0.2 (Fig.~\ref{fig1}).

\section{Accelerator experiments}
This kind of long-baseline experiments focusses on the investigation of the atmospheric
neutrino anomaly.
Typical \neu beams at accelerators are produced by protons hitting a fixed target,
where the decaying secondaries (mostly pions)
decay into \nmu{}. This dominantly \nmu beams are then used either for pure \nmu{}-\dis{}
searches or for \app{} searches by measuring electrons and/or  $\tau$-leptons produced
via charged current (CC) reactions.
The \ntau{} - \app{} search requires some beam design optimisation because the
exploration of low \delm{} values  
prefers lower
beam energies
but the $\tau$-production cross-section shows a threshold behaviour starting at
3.5 GeV and is increasing with beam energy. 
A possible \osz{} of \nmu{} into sterile \neus{} might show up in
the CC/NC ratio. 

\subsection{KEK- \sk}
\label{sect3}
The first of the accelerator based \lbls{} is the KEK-E362 experiment (K2K)
\cite{keksk} in Japan
sending a \neu{} beam from KEK to \sk. It it using two detectors, one about 300 m
away from the target and
\sk{} in a distance of about 250 km. 
The \neu{} beam is produced by 12 GeV protons from the KEK-PS hitting an Al-target
of 2cm $
\oslash \times$ 65 cm.
Using a decay tunnel
of 200 m and a magnetic horn system for focussing $\pi^+$ an almost pure
\nmu{}-beam is produced. The contamination of
\nel{} from $\mu$ and K-decay is of the order 1 \%
. The protons are extracted in a
fast extraction mode
allowing spills of a time width of 1.1 $\mu$s every 2.2 seconds. With $6 \cdot
10^{12}$ pot (protons on
target) per spill about $1 \cdot 10^{20}$ pots can be accumulated in 3 years.
The average \neu beam energy is 1.3 GeV, with a peak at about 1 GeV.
The near detector consists of two parts, a 1 kt Water-Cerenkov detector
and a fine grained
detector. The water detector is
implemented with 820 20'' PMTs and its main goal is to allow a direct comparison
with \sk events and to study
systematic effects of this detection technique.
The fine grained detector basically consists of four parts and should provide
information on the \neu beam profile 
as well as the energy distribution. First of all there are 20 layers of
scintillating fiber
trackers intersected with water. The position resolution of the fiber sheets is
about 280 $\mu$m and allows
track reconstruction of charged particles and therefore the determination of the
kinematics in the \neu{}
interaction. In addition to trigger counters there is a lead-glass counter and a
muon detector. The 
600 lead
glass counters are used for measuring electrons and therefore to determine the
\nel{} beam contamination.
The energy resolution is about 8\% /$\sqrt{E}$. The muon chambers consist of 900 drift 
tubes and 12 iron
plates. Muons generated in the water target via CC reactions can be reconstructed with a
position resolution
of 2.2 mm. The energy resolution is about 8-10 \%.
The detection method
within \sk{} is
identical to that of their atmospheric \neu detection however precise timing cuts with the beam pulse can be applied.\\
The low beam energy allows K2K only to perform a search for \nmune{} appearance and a 
general \nmu{}-disappearance. The
main background for the search
in the electron channel is quasielastic $\pi^0$-production in NC reactions, which 
can be significantly reduced
by a
cut on the electromagnetic energy. The proposed  sensitivity regions are given by \delm 
$> 2 \cdot 10^{-3} eV^2 
(3 \cdot 10^{-3} eV^2)$ and \sint $>$ 0.1 (0.4)  for \nmune{}(\nmuntau) \oszs{}, not completely covering
the atmospheric parameters.\\ 
K2K has accumulated 5.6 $\cdot 10^{19}$ pot, where the available results are based on
4.8 $\cdot 10^{19}$ pot \cite{ish}. The number of observed
events are shown in Tab.1. As can be seen, K2K observes a deficit with respect to expectation,
however the number is in good agreement with the oscillation parameters deduced from the atmospheric data.
If this $\nmu$ - \dis is becoming statistically more significant, it will be an outstanding
result.

\subsection{Fermilab-Soudan}
\label{sect4}
\begin{figure}[b]
\begin{center}
\includegraphics[width=6cm,height=5cm]{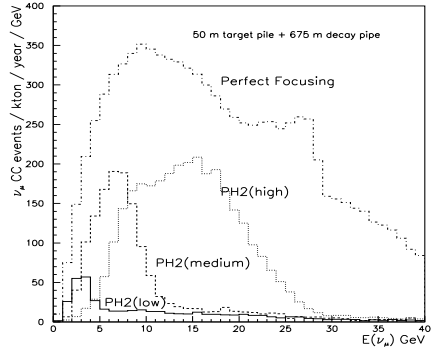}
\end{center}
\caption[]{The possible neutrino beam profiles discussed for NuMI. Motivated by
the current \delm preferred from atmospheric data, the PH2(low) beam option will 
be used.}
\label{fig2}
\end{figure}

A \neu program (NuMI) is also associated with the new Main Injector at Fermilab. The long
baseline project will
send a \neu beam produced by 120 GeV protons to the Soudan mine about 730 km away from Fermilab. Here the
MINOS experiment
\cite{minos} is under construction. It consists of a 980t near detector located at Fermilab about 900 m away 
from a graphite
target and a far detector at Soudan. The far
detector will be made of 486 magnetized iron plates, producing an average toroidal magnetic field of 
1.5 T.
They have a thickness of 2.54 cm and an octagonal shape measuring 8 m across.
They are interrupted by about 25800
m$^2$ active detector planes in form of 4.1 cm wide solid scintillator strips with x and y
readout to get the necessary tracking informations. 
Muons are identified as tracks transversing at
least 5 steel plates, with a small number of hits per plane. The total mass of the detector 
will be 5.4 kt. The neutrino beam energy can be tuned by positioning the magnetic horn
system in various positions relative to the target, resulting in different beam energies (Fig.~\ref{fig2}).
Oscillation searches in the \nmune and \nmuntau channel can be 
done in several ways. 
\nmu{} disappearance searches can be performed by investigating the visible energy distributions 
in charged current events.
A powerful way to search for oscillations is to compare the NC/CC ratio in the near and far detectors. 
By using this ratio, information on possible \nmu -
$\nu_{sterile}$ \oszs{} can be obtained, because $\nu_{sterile}$ would not contribute to
the NC rate as well.
A 10 kt$\cdot$yr exposure will cover the full atmospheric evidence region.
Start of data taking is forseen around  2005.

\subsection{CERN-Gran Sasso}
\label{sect5}
\begin{figure}[b]
\begin{center}
\includegraphics[width=7cm,height=6cm]{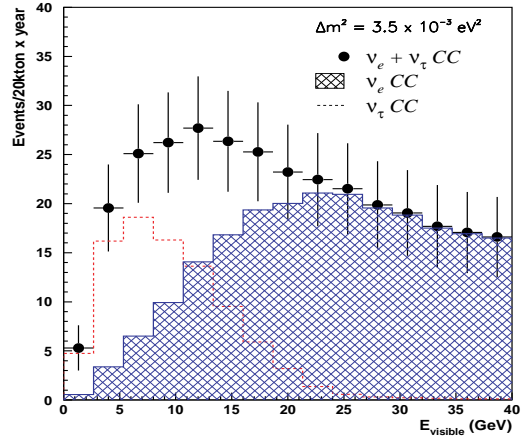}
\end{center}
\caption[]{Simulated visible energy distribution for ICARUS. The expected $\nu_e$ CC contribution is
shown as filled histogram, the possible contribution of $\nu_\tau$ CC with $\tau
\rightarrow e \nu \nu$
at low visible energy is shown as dotted line. The points represents the combined
curves including
statistical errors.}
\label{fig3}
\end{figure}

\begin{figure}[b]
\begin{center}
\includegraphics[width=8cm,height=6cm]{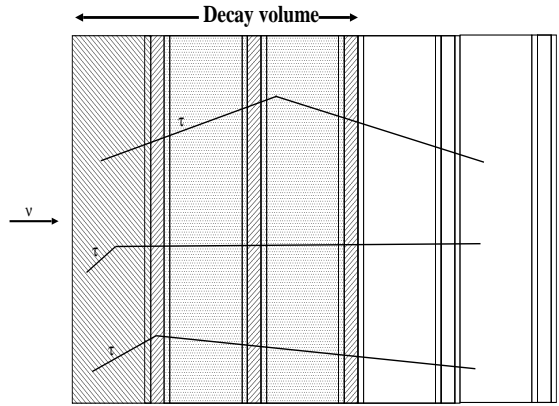}
\end{center}
\caption[]{Principal layout of an OPERA device working as an ECC. Thin emulsions sheet separated
by a small gap are intersected by lead acting as target. The three different kind of 
expected topologies for $\tau$-decays are shown.}
\label{fig4}
\end{figure}

A further program considered in Europe are \lbls{} using a \neu{} beam from CERN to
Gran Sasso Laboratory \cite{ngsprop}. The distance is 732 km. The beam protons from the SPS can 
be extracted with energies up to 400 GeV hitting a graphite target in a distance of 830 m to the SPS. 
After a magnetic horn system for focusing
a decay pipe of 1000 m will follow. The average beam energy is around 20 GeV, optimized for $\ntau$-appearance
searches.\\
Two experiments are under consideration for the Gran Sasso Laboratory to perform
an \osz search. The first proposal is the ICARUS
experiment \cite{icarus}.
This liquid Ar TPC with a modular design offers excellent energy and position 
resolution. A prototype of 600 t is
approved for installation in Gran Sasso. An update with 2 additional modules of 1200 t each is planned.
Beside a \nmu - \dis search by looking for a distortion in the energy spectra,
also an \nel - \app search can be done because of the good
electron identification capabilities.
A \ntau{}-\app{} search can be obtained will be performed by using kinematical criteria as in NOMAD.
For ICARUS a detailed analysis has been done for the
$\tau \ra e \nu \nu$ channel  (Fig.~\ref{fig3}) and is under investigation for other decay channels as well.\\
The second proposal is a \ntau - \app search with a 2 kt lead-emulsion sandwich detector
(OPERA) \cite{opera}. 
The principle idea is to use a combination of 1mm lead plates as a massive target for \neu{} interactions and two thin 
(50 $\umu$m) emulsion sheets separated by 200 $\umu$m,  
conceptually working as emulsion cloud chambers (ECC) (Fig.~\ref{fig4}).
The detector has a modular design, with a brick as basic building block, containing 56 Pb/emulsion sheets.
3264 bricks together with electronic trackers form a module. 24 modules will form a supermodule of about
652 t mass. Three supermodules interleaved with a muon spectrometer finally form the full detector.
The total number of bricks is about 235000.
The scanning of the emulsions is done by high speed automatic CCD microscopes.
The $\tau$, produced by CC reactions in the lead, can be investigated by
two signatures. For long decays the emulsion sheets are used to verify the kink of the $\tau$-decay,
while for short decays an impact parameter analysis can be performed identifying tracks not pointing
towards the primary vertex point.
The analysis here is done on an event by event basis. In five years of data
taking, corresponding to 2.25 $\cdot 10^{20}$ pot a total of 18 events are expected for  
\delm $= 3.2 \cdot 10^{-3} eV^2$.\\
Two further detectors basically designed for atmospheric neutrino detection, namely MONOLITH \cite{monolith}
and AQUA-RICH  \cite{tom} could also be envisaged to be used in accelerator based oscillation searches.

\section{Future machines}
\label{sect6}
Driven by the recent evidences for oscillations and facing the three angles and one phase in the MNS-matrix,
the idea to build new beams with very high intensity 
has been pushed forward. 
\subsection{Beta beams}
The idea is to accelerate $\beta$-unstable isotopes \cite{zuc} to energies of a few 100 MeV using ion 
accelerators like ISOLDE. This would give a clearly defined beam of \nel or \bnel. Among the favoured
isotopes discussed are $^6$He in case of a \bnel beam and $^{18}$Ne in case of a \nel beam.
\subsection{Superbeams}
Conventional neutrino beams in the GeV range run into systematics when investigating oscillations involving \nmu and \nel
because of the beam contaminations of \nel from $K_{e3}$ decays. To reduce this component, lower energy beams
with high intensity are proposed. A first realisation could be the JAERI-SK beam in Japan, in its first phase producing
a 0.77 MW beam of protons with 50 GeV on a target and using \sk as the far detector \cite{jhf}. This could be
updated in a second phase to 4 MW and also a 1 Mt detector (Hyper-K). A similar idea exists at CERN to use
the proposed SPL making a high intensity beam to Modane (130 km away). Such experiments would allow to measure $\sin^2 \theta_{23},
\delm_{23}$ and might discover $\sin^2 \theta_{13}$.
\subsection{Muon storage rings - neutrino factories}
\begin{figure}[b]
\begin{center}
\includegraphics[width=10cm,height=7cm]{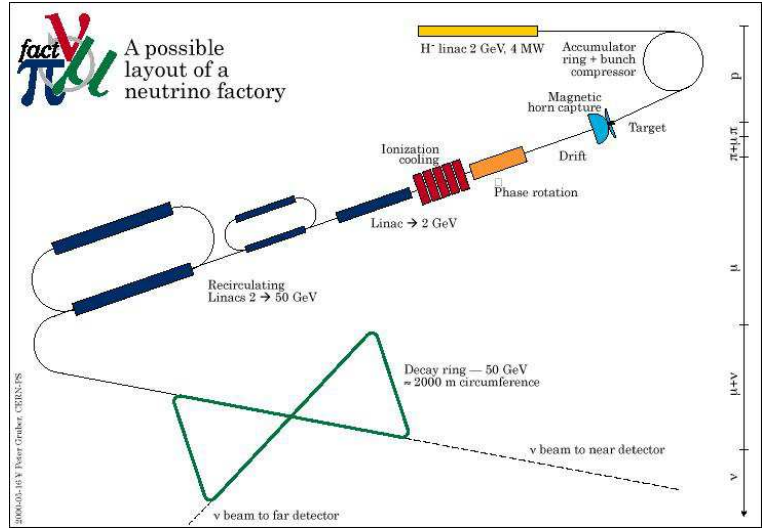}
\end{center} 
\caption[]{Conceptual design of a neutrino factory discussed at CERN. A high intensity proton beam is hitting a target to produce
secondaries. Myons from their decay will experience phase space cooling via ionisation loss, accelerated and finally brought into
a storage ring. Possible design studies are also going on at BNL, Fermilab and RAL.}
\label{fig5a}
\end{figure}

\begin{figure}[b]
\begin{center}
\includegraphics[width=6cm,height=6cm]{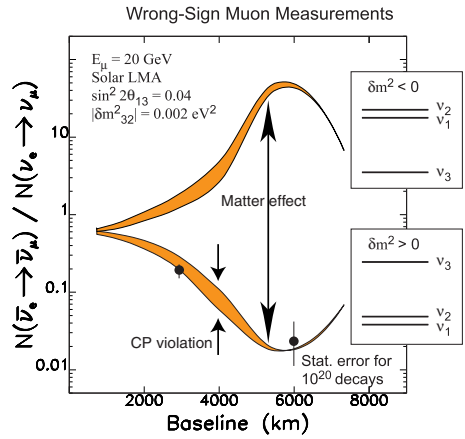}
\end{center} 
\caption[]{Ratio of the number of wrong-sign muons using $\mu^-$ and $\mu^+$ as beams. The two bands
correspond to the sign of $\Delta m_{23}^2$, the splitting shows the influence of possible matter effects
and the width represents effects of a possible CP-violating phase.}
\label{fig5}
\end{figure}

In recent years the idea to use muon storage rings to obtain high intensity neutrino beams was getting very popular
\cite{gee97}. Even many technical challenges have to be solved, it offers a unique source for future
accelerator
based neutrino physics. The two main advantages are the precisely known neutrino beam composition and the
high intensity (about 10$^{21}$ muons/year should be filled in the storage ring). A conceptional design is shown in Fig.~\ref{fig5a}.
A first experimental step towards
realisation is the HARP experiment at CERN, which will determine the target for optimal production of secondaries.
Further experimental details are under study like muon scattering (MUSCAT experiment) or muon cooling (MICE experiment),
which has to be investigated as well.
The storage ring itself could be constructed out of 2 straight sections connected by two arcs, where 
the straight regions
are used as decay regions of the muons, producing \neu beams in the corresponding directions.
Besides options using existing baselines like CERN-Gran Sasso of Fermilab-Soudan, the possibility of observing
matter effects and -- because of the strong favour of the LMA solution in solar neutrinos -- also CP violation,
it might require even longer baselines of about 3000 km (Fig.~\ref{fig5}) \cite{nufact}.

\section{Summary and conclusions}
\label{sect7}
\begin{figure}[b]
\begin{center}
\includegraphics[width=8cm,height=9.5cm]{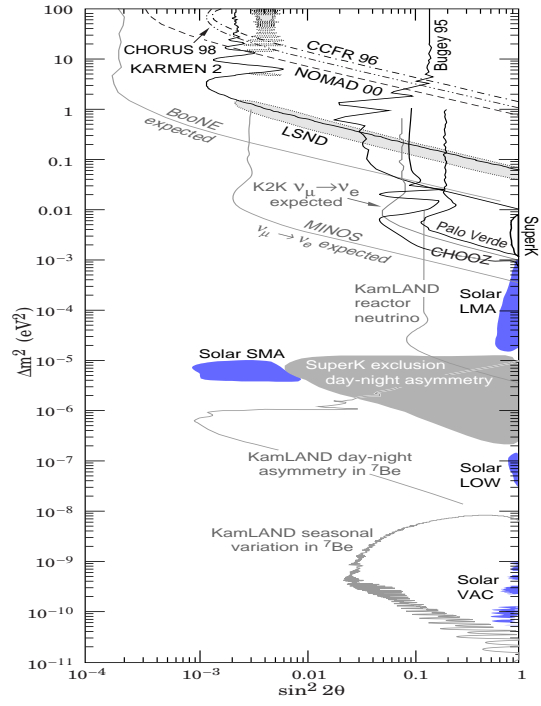}
\end{center}
\caption[]{\delm - \sint plot of all current evidences for neutrino oscillations and proposed 
goals of various future experiments.}
\label{fig6}
\end{figure}

The present evidences for \neu{} \oszs{} and their description in theoretical models requires
a variety of new \exps{} for detailed studies (Fig.~\ref{fig6}). Long baseline experiments study
the atmospheric anomaly using accelerators and the LMA solar solution with the help of
nuclear power plants. First indications of \nmu \dis are already seen in K2K, the future
activities of MINOS, ICARUS and OPERA will settle the question. KamLAND started data 
taking recently and will in the near future tell, whether LMA is indeed the correct solution
as solar neutrino data suggest.
Precise measurements of the MNS matrix elements, searching for matter effects and CP-violation
would be under study using various new high intensity beams proposed.

\begin{table}
\caption{Current results of K2K. Shown are total numbers observed within the
fiducial volume of Super-Kamiokande as well as exptected numbers for two scenarios, no oscillations
and the best fit value of atmospheric neutrino anomaly.
\sint = 1 (after
\protect
\cite{keksk})}
\begin{center}
\renewcommand{\arraystretch}{1.4}
\setlength\tabcolsep{5pt}
\begin{tabular}{cccc}
\hline\noalign{\smallskip}
Super-K&  Events & no oscillations & $\delm = 3 \cdot 10^{-3} eV^2$\\
\noalign{\smallskip}
\hline
\noalign{\smallskip}
total        & 56  & $80.6^{+7.3}_{-8.0}$  & 52.4 \\
$\mu$ -like  & 30  & 44.0 $\pm$ 6.8        & 24.4      \\
e - like     & 2   & 4.4 $\pm$ 1.7         & 3.7     \\
multi ring   & 24  & 32.2 $\pm$ 5.3        & 24.3    \\
\hline
\end{tabular}
\end{center}
\label{Tab1.1a}
\end{table}


%

\end{document}